\documentstyle[epsf,12pt]{article}         
\newcommand{\sech}{\mbox{sech}}

\title{Soliton dynamics in damped and forced  Boussinesq
equations}
\author{ E. Ar\'evalo,  Yu. Gaididei\cite{yuri} 
and F. G. Mertens\\
Physikalisches Institut, Universit\" at Bayreuth,
D95440 Bayreuth, Germany}
\date{ }
\begin{document}
\maketitle
\abstract{
We investigate the dynamics of a lattice soliton on a 
monatomic chain in the presence of damping and external forces. We
consider  Stokes and hydrodynamical damping. In
the quasi-continuum limit the discrete system leads to a
damped and forced Boussinesq equation. By using a multiple-scale
perturbation expansion up to second order in the framework of the
quasi-continuum approach  we derive a general expression for the
first-order velocity correction which improves previous results. We compare the
soliton position and shape predicted by the theory with simulations carried out on the
level of the monatomic chain system as well as on the level of the
quasi-continuum limit system. For this purpose we restrict ourselves
to specific examples, namely potentials with cubic and quartic anharmonicities as well as
the truncated Morse potential, without taking into account external forces.
For both types of damping we find a good agreement with
the numerical simulations both for the soliton position and for the
tail which appears at the rear of the soliton. Moreover we clarify why the 
quasi-continuum approximation is better in the hydrodynamical damping
case than in the Stokes damping case.\\
PACS: 63.10.+a Lattice dynamics: General theory and 05.45.Yv Solitons
}
\section{Introduction}
There is a long-standing interest in the dynamical and thermodynamical 
properties  of
anharmonic monatomic and diatomic chains ( see e.g.
 \cite{TODA81,mb84,bish,vaz,pnev,kon}).  It was shown that these
anharmonic chains can bear low-energy excitations which are
solutions of a Boussinesq  type equation (in the
long-wavelength approximation). For realistic interatomic potentials
these soliton-like excitations are supersonic and correspond to a
compression of the chain where the relation between amplitude (or
width) and velocity depends on the form of the interatomic potential.
They are very robust and propagate without energy loss, and their
collisions are almost elastic even beyond the range of validity of the
continuum approximation. Due to their robust character the soliton
excitations are important in the coherent energy transfer
and  they have been used to explain energy
transport in DNA \cite{MLC90}. There is also a growing  evidence that 
nonlinear excitations participate in the heat conduction of anisotropic 
dielectric crystals \cite{mb83,jm89,nk92,on94}. The non-diffusive heat flow was 
attributed to modified Korteweg-de-Vries solitons in \cite{on94}. The role of 
breathers for the thermal conductivity was studied in \cite{tsi}. 

So far the main attention was paid to soliton dynamics in the absence of dissipation. 
However, the dissipation influences significantly the solitons, changing their shape and 
velocity. An external driving force is therefore necessary to sustain a 
soliton in steady state. The dynamics of slowly varying solitary wave
solutions of the damped Korteweg-de-Vries equation was investigated 
 by using the methods of inverse scattering theory
 \cite{karp}, a multiple-scale perturbation expansion 
\cite{grim}, and a Green's function formalism \cite{mann}. 
The soliton motion in Toda chains  in the presence of dissipation and driving
forces was studied in \cite{hiet,kim}.

The objective of the present paper is to study properties of  solitary
waves in damped anharmonic lattices. We investigate two types of
damping: Stokes friction (in Ref. \cite{hiet} it is called outer friction)
and hydrodynamical (or internal) friction \cite{land}. We study  the case
of potentials with 
power-like anharmonicities as well as the case of a truncated
Morse potential. We compare the results of numerical simulations, which
are carried out for  discrete anharmonic lattices as well as for the
quasi-continuum Boussinesq system, with the results of a
multiple-scale perturbation expansion obtained in the framework of the
quasi-continuum approximation.

\section{System and equations of motion}
We consider a chain of equally spaced particles of mass $M~(M=1)$ with
interatomic spacing $a~(a=1)$ and displacement from equilibrium $x_n$.
The Lagrangian of our system is given by
\begin{eqnarray}\label{lagr}
L=T-U-U_{ext}.\end{eqnarray}
Here
\begin{equation}
\label{kin}
     T = \frac{1}{2} \sum_n \dot{x}^2_n(t)
\end{equation}
is the kinetic energy ( $\dot{x}\equiv\frac{dx}{dt}$),

\begin{eqnarray}\label{anh}
U=\sum_n V(x_{n+1} - x_n)\end{eqnarray}
is the  potential energy. We consider a
potential between first neighbors of two types: the power-like
potential
\begin{eqnarray}\label{power}
V(r)&=&V_{harm}+V_{anh},\nonumber\\
V_{harm}=\frac{1}{2}\,r^2,~ V_{anh}&=&\frac{1}{p}\,r^p~~~~(p=3,4,...)
\end{eqnarray}
and the truncated Morse potential
\begin{eqnarray}\label{morse}
V(r)=\frac{1}{2}\,r^2-\frac{1}{2}\,r^3+\frac{7}{24}\,r^4
\end{eqnarray}
which is the Taylor expansion of the Morse potential
$\frac{1}{2}(e^{-r}-1)^2$. The last term in Eq.
(\ref{lagr}) represents the influence of external forces and is given
by the expression
\begin{eqnarray}\label{ext}
U_{ext}=-\sum_n\,\xi_n(t)\,\left(x_{n+1}-x_n\right)
\end{eqnarray}

To take into account damping effects in soliton dynamics we introduce
the dissipation function $\Psi$\cite{land}. In the case of the Stokes friction
when the damping occurs due to interaction of the particles with a viscous
environment(outer friction) the dissipation function  depends on the velocity of 
the particles (Stokes law) and has the form 
\begin{eqnarray}\label {diss}
\Psi_s=\frac{1}{2}\,\nu\,\sum_n\,\dot{x}^2_n(t)
\end{eqnarray}
where $\nu$ is the damping constant.
In the case of internal friction (we will call this type of friction 
hydrodynamical), which is due to irreversible processes 
taking place within the system, the dissipation function  depends 
on the time derivatives of the {\it relative} displacements and is given 
by the expression 
\begin{eqnarray}\label{dish}
\Psi_h=\frac{1}{2}\,\nu\,\sum_n\left(\dot{x}_{n+1}(t)-\dot{x}_n(t)\right)^2.
\end{eqnarray}
The function (\ref{dish}) is the discrete version of the dissipation function
which is usually  used
in macroscopic elasticity  theory \cite{land}. 

It is necessary to point out that if we consider the soliton as a wave
packet, its dynamics in the presence of Stokes damping shows
that the long-wave components of the spectrum do not propagate
(App. {\bf A}). Thus the soliton decomposes after a time in the order
of $1/\nu$. This feature does not show up with the
hydrodynamical damping, if $\nu <1$ (see App. {\bf A}).

The equations of motion for the lattice displacements $x_n(t)$ in the
presence of damping have the form
\begin{eqnarray}\label{eqm}
\frac{d}{dt}\frac{\partial\,L}{\partial \dot{x}_n}-
\frac{\partial L}{\partial x_n}+\frac{\partial \Psi}{\partial
\dot{x}_n}=0
\end{eqnarray}
Substituting Eqs (\ref{lagr}) - (\ref{dish}) into Eq. (\ref{eqm}), 
the equations of motion for the relative displacement $u_n =
x_{n+1} - x_n$ can be written  as
\begin{eqnarray}
\label{eqrd}
& &\ddot{u}_n =
      V'(u_{n+1}) - 2 V' (u_n) +
     V'(u_{n-1})+\nonumber\\
& &\xi_{n+1}(t)+ \xi_{n-1}(t)-2\,\xi_n(t)+D_n 
\end{eqnarray}
where $V'(u)$ is the derivative of $V$ with respect to its
argument $u$ and the damping term $D_n$ is determined by
\begin{eqnarray}\label{damp}
D_n=\left\{\begin{array}{ll}-\nu\,\dot{u}_n& \textrm{for Stokes damping,}\\
\nu\,\left(\dot{u}_{n+1}+\dot{u}_{n-1}-2\,\dot{u}_n\right) &
\begin{array}{l} 
\textrm{for hydrodynamical} \\
\textrm{damping}
\end{array}
\end{array}\right.\nonumber
\end{eqnarray}

In order to obtain a analytical solution of the nonlinear
system of equations (\ref{eqrd}) we  apply   the quasi-continuum
approximation proposed in \cite{ros} (see also \cite{HM85,gfnm95}). 
Regarding $n$
as a continuous variable $(n \rightarrow x, u_n(t) \rightarrow
u(x,t))$,
Eq. (\ref{eqrd}) we obtain a damped and forced Boussinesq (Bq)
equation (see App. {\bf B}):
\begin{eqnarray}\label{eqc}
\partial_t^2 u-\partial^2_x u- \partial^2_t\partial_x^2 u
-\partial_x^2 \left(f(u)\right)=\,
 \nu_m\,\partial_x^m\partial_t u+\partial^2_x\xi(x,t)\nonumber\\
\end{eqnarray}
where 
$\partial _{x}$ and $\partial_{t}$ are   the derivatives with respect
to $x$ and $t$, respectively;
\begin{eqnarray}\label{force}
f(u)=\frac{d V(u)}{d u}-u
\end{eqnarray}
is  a nonlinear force and the
right-hand-side of Eq. (\ref{eqc}) represents the damping in the system and 
the action of an external force. The case $m=0$ corresponds to the
Stokes damping while the case $m=2$ corresponds to the hydrodynamical
damping: 
\begin{equation}\label{damp1}
\nu_m=\left\{\begin{array}{ll}-\nu & \textrm{if $m=0,$}\\
\nu & \textrm{if $m=2.$}\end{array}\right.
\end{equation}

\section{Multiple scale expansion}
We are interested in how the dynamics and the behavior of the
soliton is affected by the two types of  damping ($m=0,2$). So we consider
both the position of the soliton center of mass as a function
of time, $X(t)$, and its shape for
$t \,>\,0$. We make a travelling wave ansatz $u(x,t)=u(x-X(t))$ and
use a
multiple-scale perturbation expansion, developed in detail in 
App. {\bf C}, for a perturbed Bq equation
\begin{eqnarray}\label{eqc1}
\partial_t^2 u-\partial^2_x u- \partial^2_t\partial_x^2 u
-\partial_x^2 \left(f(u)\right)=\epsilon\,F(x,t)
\end{eqnarray}
where  $\epsilon\,F(x,t)$ is the  perturbation term with
\begin{eqnarray}\label{eqc2} 
F(x,t)=
\nu_m\,\partial_x^m\partial_t u+\partial^2_x\xi(x,t).
\end{eqnarray}
We seek an asymptotic solution of the form
\begin{eqnarray}\label{asym1}
u=u_0+\epsilon\,u_1+\epsilon^2\,u_2+\cdots
\end{eqnarray}
with 
\begin{eqnarray}\label{asym2}
c=c_0+\epsilon\,c_1+\epsilon^2\,c_2+\cdots
\end{eqnarray}
where $c=-\partial_t\,X(t)$ is the velocity of the soliton.
$\epsilon$ is a factor introduced for convenience in
the analytical calculations. The case $\epsilon=0$ ($u=u_0$)
reduces  Eq. (\ref{eqc1}) 
to the unperturbed Bq equation
\begin{eqnarray}\label{eqc3}
\partial_t^2 u_0-\partial^2_x u_0-\partial^2_t\partial_x^2 u_0
-\partial_x^2 \left(f(u_0)\right)=0\quad,
\end{eqnarray}
which is the well-known improved Boussinesq (IBq) equation
\cite{ros,HM85,makhankov}. When $\epsilon=1$ we recover the damped and forced Bq equation
(\ref{eqc}). In order to interpret the multiple-scale perturbation
results we must set $\epsilon=1$
and assume that the terms on the r.h.s of (\ref{eqc}) are
small enough. So, we must restrict ourselves to small values of the damping
constant $\nu$.

In what follows we restrict our study to the damped Bq equation
\begin{equation}\label{damped}
\partial_t^2 u-\partial^2_x u-\partial^2_t\partial_x^2 u
-\partial_x^2 \left(f(u)\right)=\,
 \nu_m\,\partial_x^m\partial_t u\quad.
\end{equation}
The study of the effect of external forces, particularly stochastic
forces, exceeds the frame of this paper and will be published later.

From the  multiple-scale perturbation analysis we obtain  that there
are two compatibility conditions: One of them follows from  the order
$\epsilon^{1}$ of  perturbation,
Eq. (\ref{a15}); And the other one  from the order
$\epsilon^{2}$, Eq. (\ref{a25}). Both
are valid for arbitrary potential $V(u)$.

Inserting the potential (\ref{power}) or (\ref{morse}), together with the
corresponding one soliton solution, into the 
compatibility conditions we get a set of two ordinary
differential equations of motion (ODEs). These ODEs govern the
time evolution of the  order $\epsilon^{0}$
and $\epsilon^{1}$ of velocity perturbation, namely 
$c_0$ and $c_1$, respectively. 

\subsection{The power-like anharmonic potential}
The expression  (\ref{a11}) is the one-soliton solution of the IBq equation (\ref{eqc3})
with the power-like anharmonic potential (\ref{power}).
Substituting this solution in  Eq.
(\ref{a15}) and  Eq. (\ref{a25}) yields
\begin{equation}\label{set1}
{\dot c}_0 =\left\{
\begin{array}{lr}
  - \frac{\left( p -2 \right) \, \nu \,c_0\,
       \left({c_0}^2 -1\right) }{6 - 3\,p + 
       2\,p\,{c_0}^2} &  m=0\\
          &     \\
-\frac{{\left( p-2 \right) }^2\,\nu \,
         {\left( {c_0}^2-1 \right) }^2}{
         \left( p+2 \right) \,c_0\,
         \left( 6 - 3\,p + 2\,p\,{c_0}^2 \right) } & m=2
\end{array}\right.
\end{equation}

\begin{eqnarray}\label{set2}
&{\dot c}_1&=
-\nu\frac{\left(p-2 \right) \,
       \left( 3\,\left(p-2 \right)  + 
         \left( 18 - 7\,p \right) \,{c_{0}}^2 + 
         2\,p\,{c_{0}}^4 \right) }{{\left( 6 - 3\,p + 
          2\,p\,{c_{0}}^2 \right) }^2} c_{1} \nonumber\\
& + &\,{\nu }^2
\frac{ 2\,\left(p-2 \right) \,{\sqrt{\pi }}\,c_{0}
{\Gamma(\frac{p-1}{p-2})}^2\,
     \Gamma(\frac{2 + p}{2\,p-4})}
{{\sqrt{{c_{0}}^2-1}}\,
     {\left( 6 - 3\,p + 2\,p\,{c_{0}}^2 \right) }^4\,
     \Gamma(\frac{p}{p-2})\,
     {\Gamma(\frac{p}{2\,p-4})}^2}\nonumber\\
&\times&\bigg( 6\,{\left(p-2 \right) }^3 - 
       2\,{\left(p-2 \right) }^2\,\left( -21 + 16\,p \right) \,
        {c_{0}}^2 \bigg. \nonumber\\
&+& p\,( 104 - 122\,p + 35\,p^2 ) \,
        {c_{0}}^4 + 
       p\,( 16 + 14\,p - 13\,p^2 ) \,
        {c_{0}}^6 \nonumber\\
&+&\bigg. 2\,p^3\,{c_{0}}^8 
\bigg) ~~~~\textrm{for}~~ m=0
\end{eqnarray}
and
\begin{eqnarray}\label{set3}
& &{\dot c}_1=
-\frac{\nu\,c_{1}(p-2)^2 \,(c_0^2-1)}{(p+2)^2\,c_0^2\,( 6 - 3\,p +
2\,p\,c_0^2)^2}(-3\,(p^2-4 )\nonumber\\
&-&3\,( -12 - 4\,p + p^2) \,c_0^2+ 2\,p\,(p+2) \,c_0^4)\nonumber\\
&+&\nu^2\frac{2\,(p-2 )^2\,\sqrt{\pi}\,
(c_0^2-1)^{3/2}\,{\Gamma(\frac{p-1}{p-2})}^2\,
     \Gamma(\frac{p+2}{2\,p-4})}
{( 2 + p)^2\,c_0^3\,{\left( 6 - 3\,p + 2\,p\,{c_0}^2 \right) }^4\,
     \Gamma(\frac{p}{p-2})\,
     {\Gamma(\frac{p}{2\,p-4})}^2}\nonumber\\
 &\times &\bigg(\bigg. 3\,(p-2)^4 - 3\,(p-2)^3\,(-10 + 11\,p) \,c_0^2\nonumber\\
 &+& (p-2 )^2\,p\,( -8 + 43\,p)\,c_0^4 + p\,( -32 + 84\,p - 17\,p^3) \,
 c_0^6 \nonumber\\
&+&2\,p^3\,( 6 + p ) \,c_0^8\bigg.\bigg) ~~~~\textrm{for}~~ m=2
\end{eqnarray}
where $\Gamma(\cdot)$ is the gamma function.

The set of ODEs (\ref{set1}-\ref{set3}) together with 
\begin{equation}\label{set4}
{\dot X}=c \quad\quad\textrm{where}\quad\quad
c\, =\, c_0 + c_1 
\end{equation} 
constitute the complete set of ODEs which determine $X(t)$ 
as a function of time.

\subsection{The truncated Morse potential}
The one-soliton solution of the IBq with the truncated Morse potential is
\begin{equation}\label{morse0}
u_0=\frac{A}{1+B\,\sinh^2\left(\frac{\eta}{2}\, (x-c_0\, t)\right)}
\end{equation}
where
\begin{eqnarray}
A=\frac{\pm 6(c_0^2-1)}{\mp 3+\sqrt{21 c_0^2-12}}\quad,\quad
B=\frac{2 \sqrt{21 c_0^2-12} }{\mp 3+\sqrt{21 c_0^2-12}}\nonumber
\end{eqnarray}
and
\begin{eqnarray}\label{morse1}
\eta=\frac{\sqrt{c_0^2-1}}{c_0}\quad . 
\end{eqnarray}
The upper sign means the soliton produces a rarefaction of the
lattice, and the lower sign means compression. In this
paper we consider only the compressional case.
By inserting Eq. (\ref{morse0}) in Eq. (\ref{a15}) and Eq. (\ref{a25}) we
get a set of ODEs. They are rather
cumbersome and therefore we make a further approximation by
considering only the soliton dynamics close to the sound velocity. So
we expand the ODEs in a Taylor series around the sound velocity where
$O(c_0^3)$ terms  are neglected. In the  Stokes damping case the
ODEs take the form
\begin{eqnarray}\label{morsestokes}
&{\dot c}_0&+\,\frac{68}{45}\,\nu\, c_0\,-\,\frac{19}{45}\,\nu\,
\,c_0^2\,-\,\frac{49}{45}\,\nu=0 \nonumber\\ 
&{\dot c}_1&-\frac{2\,\nu \,
     \left( -101 + 41\,c_0 \right) }{15\,
     \left( 5 + 7\,c_0 \right)}\,c_1\nonumber\\ 
&-& \frac{2 \,{\nu }^2\,{\sqrt{2}}\,
     \left( -39 - 1112\,c_0 + 
       851\,{c_0}^2 \right) }{225\,
     {\sqrt{c_0-1}}\,
     \left( 5 + 7\,c_0 \right) }=0, 
\end{eqnarray}
and in the hydrodynamical damping case ($m=2$) 
\begin{eqnarray}\label{morsehydro}
&{\dot c}_0&
 -\frac{8}{15} \,\nu \,c_0 + 
 \frac{4 }{15} \,\nu\,{c_0}^2+ \frac{4 }{15}\,\nu
  =0\nonumber\\ 
&{\dot c}_1&+
\frac{32\,\nu\,
     \left(c_0 -1 \right) }{5\,
     \left( 5 + 7\,c_0 \right)} \,c_1 - 
  \frac{176\,{\sqrt{2}}\,{\nu }^2\,
     {\left( c_0 -1 \right) }^{\frac{3}{2}}}{75\,
     \left( 5 + 7\,c_0 \right) } =0\quad .
\end{eqnarray}
These approximations of the
exact equations are good within a range of velocities
$1<\,c_0\,\; {}^{<}_{\sim}\; 1.1\quad$\\

Notice that  either Eqs. (\ref{morsestokes}) or Eqs. (\ref{morsehydro}), 
together with  Eq. (\ref{set4}), constitute a complete set of
equations of motion for $X(t)$.

\section{Numerical simulations}
In order to verify these  theoretical results for both the power-like
and the Morse potentials we have performed molecular dynamics
simulation  for the discrete monatomic chain  which is
governed by Eq. (\ref{eqrd}). Moreover we have performed simulations
on the level of the quasi-continuum limit, namely with the damped
Bq equation (\ref{damped}).
For the damped Bq system we have used finite-difference discretization in the
space-domain \cite{Bogolubsky} (see App. {\bf D} for details). The
time integration in 
both kinds of simulations was carried out by using the Heun method \cite{Heun}.
In order to start the simulations at $t=0$ we have used one-soliton
solutions of the  IBq equation. Since for some cases we need a
long simulation time and the solitons are supersonic, we have used
periodic boundary conditions to reduce the size of our system. In
fact, we have used a chain with 1500 lattice points in the molecular
dynamics simulation; and in the damped-Bq-simulation  the length of the
system has been $L=1000$ with $\Delta x = 0.25$. We remark here
that the solitons are bounded even when they develop
a tail in the presence of perturbations, because the rear of this tail
vanish. As the soliton including its tail is bounded in our
finite system we are allowed to use periodic boundaries in our
codes. The length of the tail grows with time, so we have considered
not too long simulation times to avoid a possible overlapping between
the rear of the tail and the front of the
soliton. The other parameters have been
$\Delta t = 10^{-2}$ in the molecular dynamics simulation and 
$\Delta t = 10^{-1}$ in the damped-Bq-simulation.

We have checked the accuracy of
our codes by calculating the conserved quantity
\begin{displaymath}\nonumber
\int_{-\infty}^{\infty}u(x,t)\,dx 
\end{displaymath} 
which is valid not only for the free soliton case ($\nu_m=0$) but also for the
hydrodynamical damping case ($m=2$). For the longest simulation
time the variation of this conserved quantity has been lower than 
$4\times 10^{-9}\%$ in the molecular dynamics simulation 
and lower than $2\times 10^{-13}\%$ in the
damped-Bq-simulation. Notice that this conserved quantity can be calculated in
a numerical window  as long as there is not overlap between its tail
and the front of the soliton.

The center of the soliton in both types of simulation has been found by finding
the three points  $x_{i-1}$, $x_i$ and $x_{i+1}$ where
$u(x_i)$ is the absolute discrete  maximum or minimum depending on whether the
soliton is rarefactive or compressional.
Afterwards a parabola has been fitted to the three coordinates, namely $\{x_i,u(x_i)\}$,
and we have defined the vertex of this parabola as the soliton
center of mass.

We have chosen $\nu_0\,=\,-\nu\,=-\,10^{-3}$ and
$\nu_2\,=\,\nu\,=\,10^{-2}$. The reason for choosing different values
is that the Stokes damping has a stronger effect than the
hydrodynamical damping for the same value $\nu$.\\

\subsection{Soliton dynamics in the presence of  hydrodynamical
damping ($m=2$)} 

\begin{figure}
\centerline{\epsfxsize=9.0truecm \epsffile{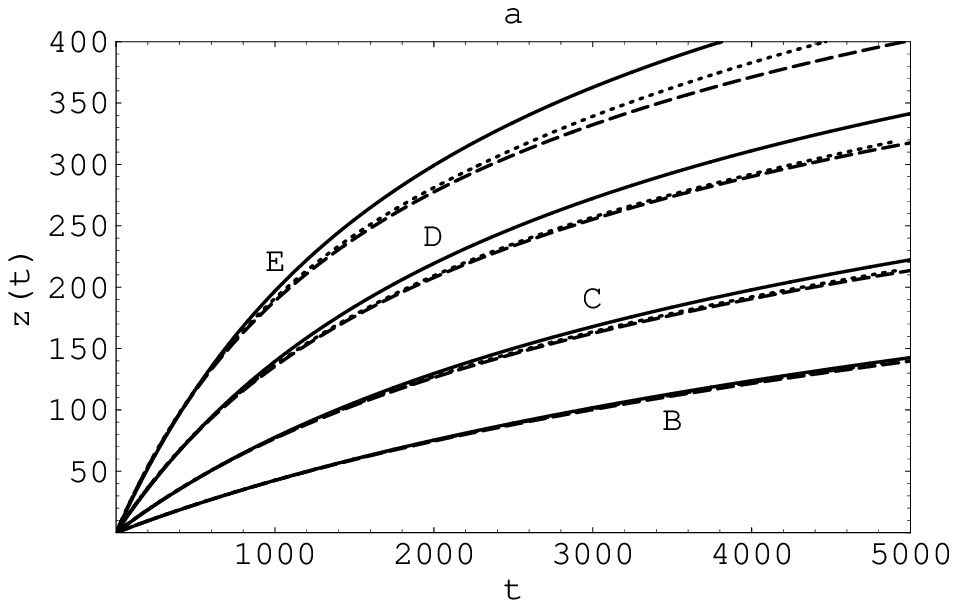}}
\centerline{\epsfxsize=9.0truecm \epsffile{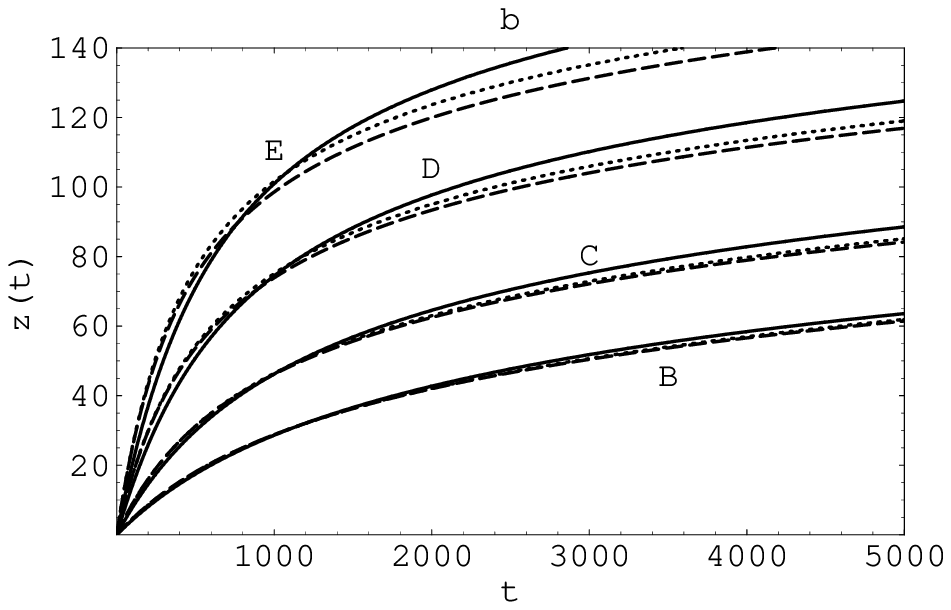}}
\centerline{\epsfxsize=9.0truecm \epsffile{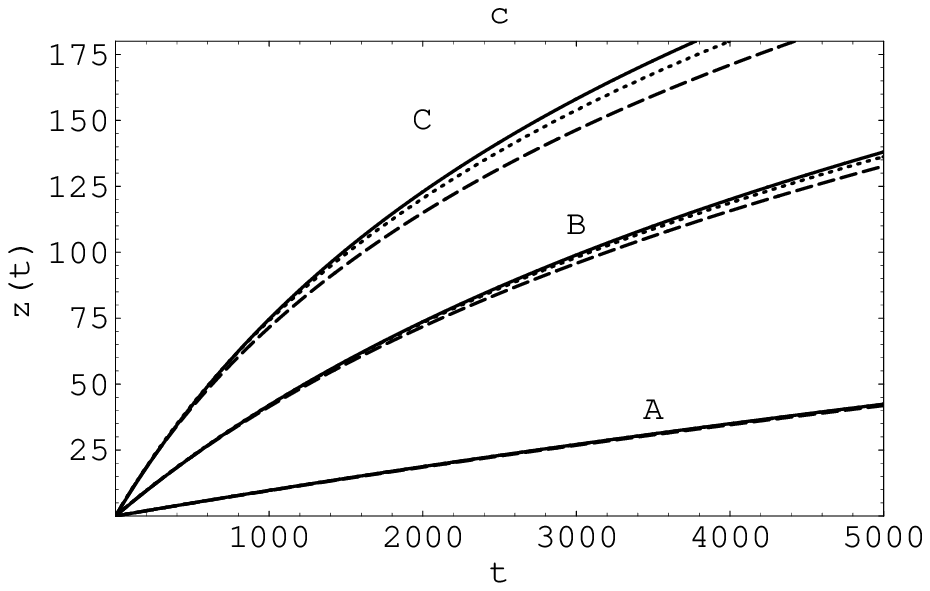}}
\caption{Soliton position in the sound velocity moving frame $z$ vs. time in
the hydrodynamical damping case with $\nu\,=\,10^{-2}$. Solid line: molecular dynamics
simulation, dashed line:  theory's prediction (solution of
Eqs.(\ref{setm2p3}-\ref{setm2p4})), dotted line: simulation for the Bq
equation (\ref{damped}). Fig. (a) and (b) correspond to the cubic
and quartic anharmonicities, respectively, and Fig. (c)
corresponds to the truncated Morse potential. The uppercase letters A,
B, C, D, and E correspond to different initial velocities,  
namely $\,c(0)=1.01$, $ 1.05$, $1.1$, $1.2$ and  $1.3$, respectively}
\label{fig1}
\end{figure}

In the case of the power-like anharmonic potential we have performed
simulations in specific cases, namely  for cubic  as well as  for
quartic anharmonicity. We have also performed simulations in the case
of the truncated Morse potential. 

In the cubic case  ($p=3$) in the presence of hydrodynamical damping 
Eqs.(\ref{set1}-\ref{set3}) reduce to
\begin{eqnarray}\label{setm2p3}
&{\dot c}_0&+\frac{ \nu \,{\left( {c_0}^2 - 1 \right) }^
          2  }{15\,\left( -c_0 +
         2\,{c_0}^3 \right) }=0\nonumber\\
&{\dot c}_1&
-\,\nu \frac{{\eta }^2\,\left( -4 - 3\,{\eta }^2 + 2\,{\eta }^4 + 
       {\eta }^6 \right)}{15\,
     {\left( 1 + {\eta }^2 \right) }^3} \,c_1 \nonumber\\
& &-{\nu }^2 \frac{{\eta }^3\,\left( 22 + 40\,{\eta }^2 + 18\,{\eta }^4 + 
         {\eta }^6 \right) }{225\,
     {\left( 1 + {\eta }^2 \right) }^3}=0\nonumber\\
\end{eqnarray}
where $\eta$ has been defined in (\ref{morse1}).
In same way, for the quartic case ($p=4$) Eqs.
(\ref{set1}-\ref{set3}) reduce to 
\begin{eqnarray}\label{setm2p4}
&{\dot c}_0&+{\frac{ \nu \,{\left( {c_0}^2 - 1 \right) }^
          2 }{3\,\left( -3\,c_0 +
         4\,{c_0}^3 \right) }}=0\nonumber\\
&{\dot c}_1&- \,\nu \frac{{\eta }^3\,\left( -4 - 3\,{\eta }^2 +
3\,{\eta }^4 \right)}{3\,{\left( 1 + 3\,{\eta }^2 \right) }^2}\,c_1 \nonumber\\
& &-{\nu }^2\frac{{\pi }^2\,{\eta }^5\,
\left( 9 + 29\,{\eta }^2 + 39\,{\eta }^4 + 3\,{\eta }^6 \right)}
{36\,{\left( 1 + 3\,{\eta }^2 \right) }^4}
=0. \nonumber\\
\end{eqnarray}
In the case of the truncated Morse potential, which contains 
a combination of cubic and quartic anharmonicities,
we have already got the corresponding set of simplified
Eqs. (\ref{morsehydro}). So depending on the type of anharmonicity  we
have solved either Eqs. (\ref{setm2p3}) or
Eqs. (\ref{setm2p4}), or Eqs. (\ref{morsehydro}) together  with
Eq. (\ref{set4}). The result from those numerical solutions is what we
call theory's prediction.

In Fig. \ref{fig1} we show several examples of the soliton position as a
function of time. These examples follow from the two kinds of
simulations and from theory's prediction. In particular
Fig. \ref{fig1}a and  Fig. \ref{fig1}b correspond to the 
soliton dynamics of the cubic and
quartic cases, respectively. And  Fig. \ref{fig1}c
corresponds to the case of the truncated Morse
potential. In each figure we show
the soliton position for different initial velocities. These cases are
denoted by uppercase letters. The position is plotted in the
sound velocity moving frame, defined by $\textrm{z}(t)=X(t)-t$.
Notice that we consider lower initial velocities in the case of the
truncated Morse potential (Fig. \ref{fig1}c), because in deriving
Eqs. (\ref{morsehydro}) further approximations have been made.
In general for all the potentials we see that for low initial
velocities (cases A, B and C) the lattice soliton position
(solid-lines) is predicted rather well by theory's prediction
(dashed lines) as well as by the position of the Bq-soliton (dotted
lines). For higher initial velocities, namely $c(0)=1.2$ (cases D),
the position of the lattice soliton agrees better with the
position of the Bq soliton (dotted lines) than with 
the theory (dashed lines). And for even higher initial velocities,
namely $c(0)=1.3$ (cases E), quantitatively there is a clear
difference between the three dynamics. However qualitatively they are
similar. The reason for this behavior is that the quasi-continuum approximation
naturally predicts better the dynamics of broad solitons than that of 
narrow ones \cite{HM85}; and the solitons are narrower when
the initial velocity is higher. Notice that in the cubic and quartic
cases  theory(dashed lines) predicts better the Bq soliton
position (dotted lines) than the lattice soliton position
(solid-lines). This is due to the fact that both the theory and the Bq
equation have been derived in the framework of the quasi-continuum
limit. This feature does not show up in the case of the truncated
Morse potential due to the further approximations that we have made in
the theory of this case. 

\subsection{Soliton dynamics in the presence of Stokes damping ($m=0$)}

In this section we treat the same cases as in the previous section but
with the soliton bearing systems in the presence of Stokes
damping.

As in the previous section  Eqs. (\ref{set1}-\ref{set3}) can be
reduced depending on whether the potential has  cubic or quartic anharmonicity. In
particular for the cubic anharmonicity ($p=3$)
\begin{eqnarray}\label{setm0p3}
&{\dot c}_0&+\frac{\nu\,c_0\,
\left( {c_0}^2 - 1 \right) }{6\,{c_0}^2-3}=0 \nonumber\\
&{\dot c}_1&
+\nu \frac{\left( 2 + {\eta }^2 + {\eta }^6 \right)}
{3\,{\left( 1 + {\eta }^2 \right) }^3}\,c_1
-{\nu }^2\frac{\left( -4 + 3\,{\eta }^2 + 8\,{\eta }^4 + 2\,{\eta }^6 
\right) }{9\,\eta \,{\left( 1 + {\eta }^2 \right) }^3} 
=0.\nonumber\\
\end{eqnarray}
In the same way the reduced set of equations for the  quartic case
is
\begin{eqnarray}\label{setm0p4}
&{\dot c}_0&+{\frac{\nu \,c_0\,
       \left( {c_0}^2 -1 \right) }{
       4\,{c_0}^2 - 3}}=0\nonumber\\
&{\dot c}_1&+ \,\nu\frac{{\left( \eta  + 3\,{\eta }^3 \right) }^2\,
     \left( 2 - {\eta }^2 + 3\,{\eta }^4 \right)}
{{\eta }^2\,{\left( 1 + 3\,{\eta }^2 \right) }^4}\,c_1- \nonumber\\
& & ~~~~\,{\nu }^2\frac{ {\pi }^2\,\left( -1 - 3\,{\eta }^2 - 5\,{\eta }^4 + 
         19\,{\eta }^6 + 6\,{\eta }^8 \right) }{4\,
     \eta \,{\left( 1 + 3\,{\eta }^2 \right) }^4}
=0\nonumber\\
\end{eqnarray}
And Eqs. (\ref{morsehydro}) correspond to the case of the truncated Morse
potential. 

The theory's prediction of the soliton position follows
from the numerical solution of either Eqs.(\ref{setm0p3}) or
Eqs.(\ref{setm0p4}) or Eqs. (\ref{morsestokes}) together with
Eq. (\ref{set4}). 
\begin{figure}
\centerline{\epsfxsize=9.0truecm \epsffile{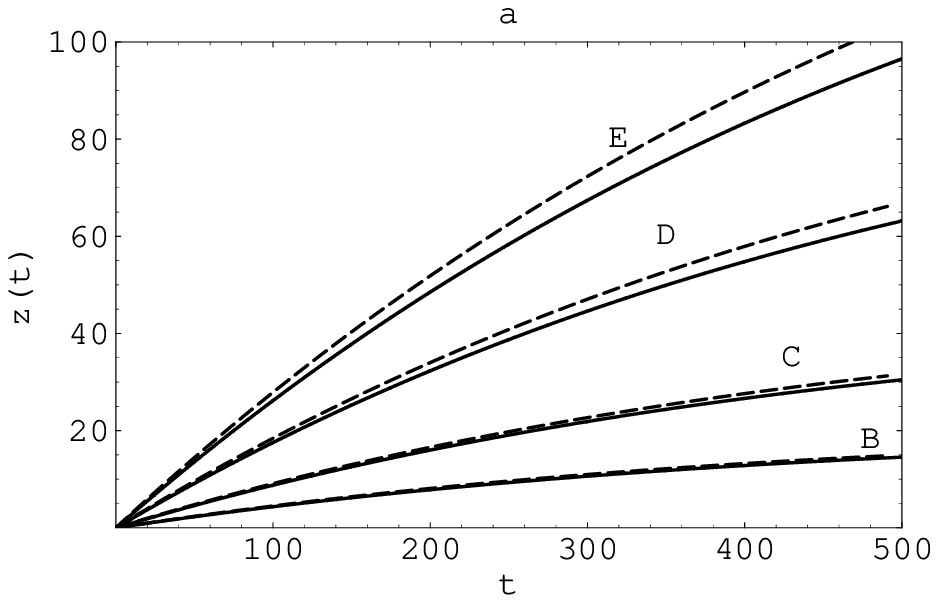}}
\centerline{\epsfxsize=9.0truecm \epsffile{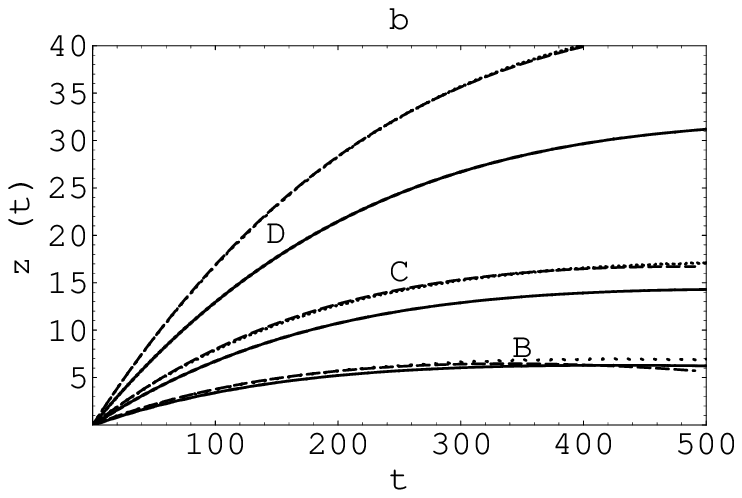}}
\centerline{\epsfxsize=9.0truecm \epsffile{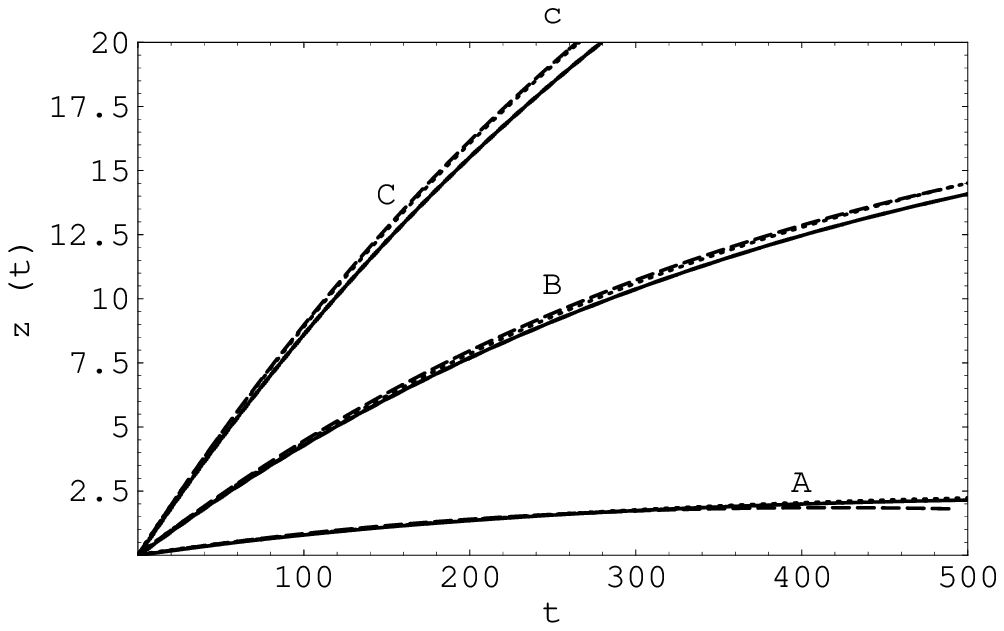}}
\caption{Soliton position in the sound velocity moving frame $z(t)$
vs. time in the Stokes  damping case with $\nu\,=\,10^{-3}$. Solid
line: molecular dynamics 
simulation, dashed line:  theoretical prediction (solution of
Eqs.(\ref{setm0p3}-\ref{setm0p4})), dotted line: simulation of the Bq
equation (\ref{damped}). Fig. (a) and Fig. (b) correspond to the cubic
and quartic anharmonicities respectively. And Fig. (c)
corresponds to the truncated Morse potential.
The uppercase letters A, B, C, D, and E correspond to different
initial velocities, namely $\,c(0)=1.01$, $ 1.05$, $1.1$, $1.2$ and
$1.3$, respectively}  
\label{fig2}
\end{figure}

We show in Fig. \ref{fig2} the same cases that we have treated in Fig.
\ref{fig1}. We have also kept the same convention. 
In Fig.
\ref{fig2}a we have not plotted the position of the Bq soliton because it
agrees very well with the  theoretical prediction (dashed line). In
general we can extend the comments made for the hydrodynamical damping
case to the present Stokes damping case. 
We only want to remark some relevant differences. First, in the present case
we have plotted the soliton dynamics during the transient regime of
the system, namely $\,t\,<\,1/\nu\,$, in contrast to the
hydrodynamical damping case where we have also considered times
$\,t\,>>\,1/\nu\,$. It is because the overdamp character of the Stokes
damping, in fact, the lattice soliton is destroyed by the damping, namely
for times $t\,{}^{>}_{\sim}3/\nu$. On the other hand, neither the Bq
simulation nor the analytical results predict in a correct way
the lattice soliton 
dynamics for times $\,t\,{}^{>}_{\sim}\,1/\nu\,$.
And second, the agreement between the position of
the lattice soliton (solid-line) and either the
position of the Bq soliton (dotted line) or theory's prediction
(dashed line) is not as good as in the hydrodymamical case. For instance,
if we compare the results for the quartic case with $c_0(0)=1.2$ (Figs. \ref{fig1}b and
\ref{fig2}b: case D) we see that the agreement between simulations and
theory is better for the hydrodynamical damping case than for the Stokes
damping case. 

\subsection{Soliton profile.}

\begin{figure}
\centerline{\epsfxsize=9.0truecm \epsffile{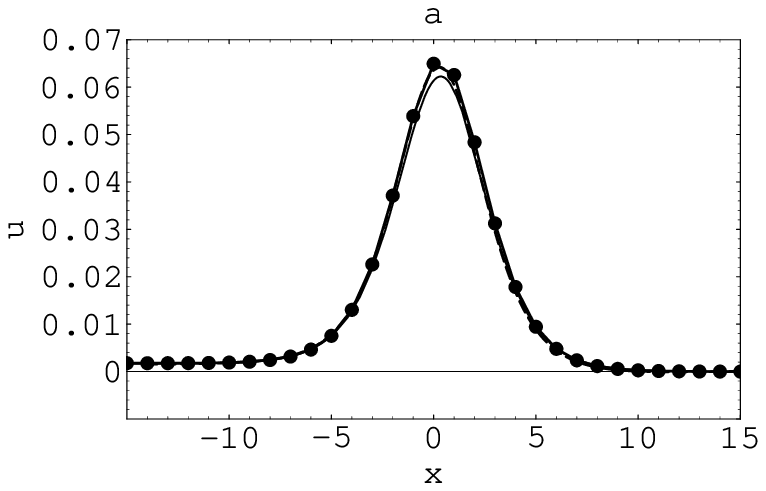}}
\centerline{\epsfxsize=9.0truecm \epsffile{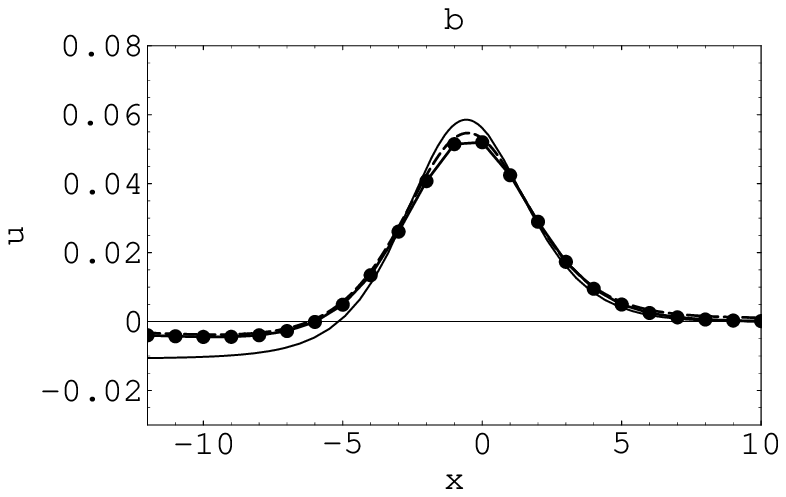}}
\caption{Snapshots of the soliton profile for the system with cubic
anharmonicity. a: hydrodynamical damping case at $t=5000$ with $c(0)=1.05$,
and $\nu=10^{-2}$. b: Stokes damping case at $\,t=500$ with $c(0)=1.1$
and  $\nu=10^{-3}$. Solid circles: lattice soliton profile, dashed lines:
Bq soliton profile, thin solid lines: theory's prediction.}
\label{fig3}
\end{figure}

Up to now we have analyzed the dynamics of solitons but
not their shape under the influence of damping.
The soliton profile for  $t>0$ in the presence
of either hydrodynamical damping or Stokes damping can be
obtained by a multiple-scale perturbation theory (App. {\bf C}).
In first order the soliton solution reads
\begin{equation}
u=u_0+u_1
\end{equation}
where $u_0$ is the unperturbed solution (\ref{a11}) and
the function $u_1$ follows from Eqs. (\ref{a17}) and (\ref{a}).
As an example, the soliton solution in 
the case of a cubic potential with hydrodynamical damping reads
\begin{equation}\label{solsol1}
u=u_0+\frac{1}{2} M + w + v \nonumber \\
\end{equation}
with
\begin{eqnarray}\label{solsol2}
w&=&\sech^2(\phi)\,(A_1+A_2\, \phi\, \tanh(\phi)) \nonumber\\
v&=&
A_3\,\phi\,\sech^2(\phi)+(A_4+(A_5\,\phi^2+A_6)\sech^2(\phi))\,\tanh(\phi)\nonumber\\
&+&A_7\,Tanh^3(\phi)
\nonumber
\end{eqnarray}
where $\phi=\eta\theta/2$.  $M$ and the coefficients $A_i,\, i=1,..,7$ depend only
on the time and are written down explicitly in App. {\bf E}. In
order to have a look at this 
theoretical behavior compared with the results from the
simulations we show in Fig. \ref{fig3} two specific examples which
belong to the cubic case. Figs. \ref{fig3}a and  \ref{fig3}b show
snapshots of the soliton profile moving to the right side in the presence of  
hydrodynamical damping and Stokes damping, respectively. 
We see that in both examples
the profiles are asymmetric and agree with each other rather well.
The main feature which differs in both figures is that the amplitude of
the tail which appears at the rear of the soliton in the presence of
the hydrodynamical damping is positive, while in the other case it is
negative. Notice that the theory's prediction  of the Stokes damping
case is not as good as in the hydrodynamical case where the
dieviations are very small, in fact they are visible only in the
center of the soliton.

\section{Discussion and conclusions}

In summary, in this work we have developed an analytical theory for the
dynamics of lattice solitons on a monatomic chain under the 
influence of damping. We have considered Stokes and hydrodynamical
damping.  In the quasi-continuum approximation the
dynamics of the driven and damped anharmonic lattice has been described by
a driven and damped Boussinesq equation. Our analytical approach has
been based on the multiple-scale perturbation theory.  We have derived sets of
equations of motion corresponding to zero- and first order
perturbations for the 
velocity. We have also calculated the first order perturbation for
the soliton profile which develops a tail at the rear end. In order to
check the validity of our results we 
have performed molecular dynamics simulation of the 
damped anharmonic lattice. We have also solved numerically the
damped Bq equation. We have considered lattices with cubic and
quartic anharmonicities. The soliton position
has been defined as the position of the maximum of the soliton.
We have observed that our theory predicts in a correct way 
the dynamics of  lattice solitons when they  propagate  in a medium
with either hydrodynamical or Stokes damping. This good agreement also holds for the 
soliton profiles. However, in the Stokes damping case our analysis is
only done for the transient time, namely $t < 1/\nu$ where $\nu$ is
the damping constant, because the soliton decomposes for larger times.

We have noticed that the quasi-continuum approximation
describes in a better way the dynamics of the lattice solitons in the
presence of hydrodynamical damping than in the case of the Stokes
damping. This difference is due to the fact that in the hydrodynamical damping
case the long-wave linear modes, which mostly contribute to the soliton dynamics,
are underdamped (see Eq.(\ref{underdamped})) while in the
other case they are overdamped.  

In general, the agreement between our theory and molecular
dynamics simulation is mostly due to fact that our theory has been
derived in the framework  of the quasi-continuum limit. Moreover, this
approach is better than earlier  approximations made for the
Korteweg-de Vries equation \cite{grim}, because 
higher soliton velocities can be considered.

{\bf Acknowledgements}

We acknowledge support from DLR grant Nr.: UKR-002-99.
Yu. Gaididei is grateful for the hospitality of the University of Bayreuth 
where this work was performed. 

\newpage
\appendix

\section{Damping}
The Stokes damping does not permit the long wave components of  a wave
packet to propagate, while the hydrodynamical damping under a
certain condition does not show  this feature.  This can be seen by
means of a simple example: let us consider a harmonic monatomic chain with $2N$
lattice points whose equations of motion 
for the relative displacements $u_n(t)$ in  the presence of Stokes
damping have the form
\begin{eqnarray}\label{apa4}
\ddot{u}_n(t)=u_{n+1}(t)-2u_n(t)+u_{n-1}(t)-\nu  \dot{u}_n(t)
\end{eqnarray}
with $n=1,2,3,...,2N-1$.
A  travelling wave packet may be written as
\begin{equation}\label{apa6}
u_n(t)=\sum_{k=0}^{2N-1}\tilde u_k e^{-i(\beta_k n-\omega_k t)}
\end{equation}
where $\beta_k$ is the wave number.

By inserting (\ref{apa6}) in (\ref{apa4}) we get $2N$  equations of
motion in  $k$-space:
\begin{equation}\label{apa11}
\omega_k^2-i\nu  \omega_k-\tilde{\gamma}_k=0
\end{equation}
where
\begin{equation}\label{ap13}
\tilde{\gamma }_k=2\left(1-\cos(\beta_k)\right)
\end{equation}
is a $k$-dependent function which satisfies
\begin{equation}\label{gammacondition}
0\leq \tilde{\gamma }_k\leq 4\quad.
\end{equation}
Solving  (\ref{apa11}) we obtain
\begin{equation}\label{overdamped}
i\omega_k=-\frac{\nu }{2}\pm
 i\sqrt{\tilde \gamma_k-\left(\frac{\nu }{2}\right)^2}\quad .
\end{equation}
Notice that for every finite value of $\nu  >0$ there are small values
of $k$ for which the condition of oscillation
\begin{equation}
\tilde\gamma_k-\left(\frac{\nu }{2}\right)^2 > 0
\end{equation}
is not satisfied. In other words, parts of the wave packet do not
propagate. 

However, this is not the case for the
hydrodynamical damping. Here the equations of motion are
\begin{eqnarray}\label{apa14}
\ddot{u}_n(t)&=&u_{n+1}(t)-2u_n(t)+u_{n-1}(t)\nonumber\\
& &+\nu \left(\dot{u}_{n+1}(t)-2 \dot{u}_n(t)+\dot{u}_{n-1}(t)\right)
\end{eqnarray}
with\quad $n=1,2,3,...,2N-1 $
By using the ansatz (\ref{apa6}) 
this set of equations reads in  $k$-space
\begin{equation}
\omega_k^2 -i\nu  \tilde\gamma_k\omega_k-\tilde\gamma_k=0
\end{equation}
with $k=1,2,3,...,2N-1$.
\begin{equation}\label{underdamped}
i\omega_k=-\frac{\nu }{2}\tilde\gamma_k\pm 
i\sqrt{\tilde\gamma_k\left(1-\left(\frac{\nu }{2}\right)^2\tilde\gamma_k\right)}\quad .
\end{equation}
Here the condition of oscillation 
\begin{equation}
1-\left(\frac{\nu }{2}\right)^2\tilde\gamma_k >0
\end{equation}
is always satisfied, if $\nu  < 1$ .

\section{Quasi-continuum approximation}
The equations of motion of the lattice in the presence of
dissipation and external forces are given by Eqs. (\ref{eqrd}). 
In order to compact the calculations we consider here a dissipation
term  $D_n$ containing both the Stokes
and the hydrodynamical damping. So in this case the equations of
motion read 
\begin{eqnarray}
& &\ddot{u}_n =
      V'(u_{n+1}) - 2 V' (u_n) +
     V'(u_{n-1})+\nonumber\\
& &\xi_{n+1}(t)+ \xi_{n-1}(t)-2\,\xi_n(t)+\nu_{0}\dot{u}_n+\nonumber\\
& &\nu_{2}(\dot{u}_{n+1}-2\dot{u}_n+\dot{u}_{n-1}).
\end{eqnarray}
where $\nu_{m}$, defined by
Eq. (\ref{damp1}), with $m\,=0,\,2$ are the damping constants of 
Stokes and hydrodynamical damping, respectively.
This equation can be rewritten as
\begin{eqnarray}\label{qcaapp1}
& &\ddot{u}_n =
\hat{\gamma}(n)\bigg(V' (u_n) +\xi_n(t)+\nu_{2}\dot{u}_n\bigg)+\nu_{0}\dot{u}_n
\end{eqnarray}
where 
\begin{equation}
\hat{\gamma}(n)=4\sinh^2\left(\frac{\partial_n}{2}\right)=4\sinh^2\left(\frac{a}{2}\partial_x\right)
\end{equation}
is a differential operator where $x=n\,a$ and  $a$ is the lattice constant. 
At this point $x$ is regarded as a continuous variable, so
$u_n(t)\rightarrow u(x,t)$ and $\xi_n(t)\rightarrow \xi(x,t)$.
Taking into account that the function $4\sinh^2(d/2)/d^2$ is smooth at
$d=0$, we can multiply both sides of (\ref{qcaapp1}) by the operator
$a^2\partial_x^2/4\sinh^2(a\partial_x/2)$, and expanding this operator as
well as the operator on the r.h.s. in a Taylor series we get
\begin{eqnarray}\label{with}
\partial_t^2u(x,t)=a^2\partial_x^2V'+a^2\partial_x^2\xi(x,t)
+a^2\lambda\partial_x^2\partial_t^2u(x,t)\nonumber\\
+\nu_{0}\partial_t u(x,t)
-\nu_{0}a^2\lambda\partial_x^2\partial_tu(x,t)+
\nu_{2}a^2\partial_x^2\partial_tu(x,t).\nonumber\\
\end{eqnarray}
Setting $a=1$, scaling  $x\rightarrow \sqrt{\lambda}\,x$, $t\rightarrow \sqrt{\lambda}\,t$,
$\nu_{0}\rightarrow \nu_{0}/\sqrt{\lambda}$, $\nu_{2}\rightarrow
\sqrt{\lambda}\,\nu_{2}$ 
and using the definition (\ref{force}) we get
\begin{eqnarray}\label{with1}
& &\partial_t^2u(x,t)-\partial_x^2u(x,t)-\partial_x^2f(u(x,t))
-\partial_x^2\partial_t^2u(x,t)=\nonumber\\
& &+\partial_x^2\xi(x,t)+\nu_{0}\partial_t u(x,t)
-\nu_{0}\partial_x^2\partial_tu(x,t)\nonumber\\
& &+\nu_{2}\partial_x^2\partial_tu(x,t).
\end{eqnarray}
One of the Stokes damping terms can be neglected because
the field $u(x,t)$ is slowly varying in space, therefore
\begin{equation}\label{estimate}
\vert \nu_{0}\,\partial_t u(x,t) \vert >> \vert
\nu_{0}\,\partial_x^2\partial_tu(x,t)\vert.
\end{equation}
The estimate (\ref{estimate}) has been
confirmed by the numerical solution of Eq.(\ref{with}) with and without
the term  $\nu_{0}\partial_x^2\partial_tu(x,t)$. In the rest of paper
we consider separately either the Stokes damping case or the
hydrodynamical case, therefore  Eq.(\ref{with1}) can 
be written as
\begin{eqnarray}
\partial_t^2 u-\partial^2_x u- \partial^2_t\partial_x^2 u
-\partial_x^2 \left(f(u)\right)=\,
 \nu_m\,\partial_x^m\partial_t u+\partial^2_x\xi(x,t) \quad \nonumber
\end{eqnarray}
with $m=0,2$.

\section{Multiple-scale perturbation expansion \label{app-aprox}}
In this appendix we develop a multiple-scale perturbation approach to the
generalized Boussinesq-Burgers equation
\begin{eqnarray}
\label{a1}
& &\partial_t^2 u-\partial^2_x u-\partial^2_t\partial_x^2 u
-\partial_x^2 \left(f(u)\right)=\nonumber\\
& &\quad\quad\epsilon\,
\left( \nu_m\,\partial_x^m\partial_t u+\partial^2_x\xi(x,t)\right)
\end{eqnarray}
where $f(u)=\frac{d V(u)}{d u}-u$ is  a nonlinear force and the
right-hand-side of this equation represents the damping in the system and 
the action of an external force.
We consider two types of damping
\begin{displaymath}
\nu_m=\left\{\begin{array}{ll}-\nu & \textrm{if $m=0$}\\
\nu & \textrm{if $m=2$}\end{array}\right.
\end{displaymath}
and $\epsilon$ is a small parameter.
In our derivation we will follow the procedure which was proposed in
\cite{grim} for the perturbed Korteweg-de-Vries equation.

By using the transformation to the moving frame of reference
\begin{eqnarray}\label{a2}
\theta = x-X(T),~~~T=\epsilon t,\nonumber\\
X(T)=\frac{1}{\epsilon}\int\limits_0^T\,c(T')dT'
\end{eqnarray}
 where $X(T)$ is the center of mass position of the soliton and 
 $c(T)$ is its velocity which depend on the "slow" time variable $T$,
 Eq. (\ref{a1}) can be written in the form
 \begin{eqnarray}
 \label{a3}
& & \left(c^2\,\partial_\theta^2-2 \,\epsilon\, c\,
 \partial_\theta\,\partial_T-\epsilon 
 \dot{c}\,\partial_\theta+\epsilon^2\,\partial_T^2\right)
 \,(1-\partial^2_\theta)\,u\nonumber\\
& &-\partial^2_\theta
u-\partial^2_\theta\left(f(u)\right)=
\epsilon\,\left(\nu_m\,\partial^m_\theta
\,\left(\epsilon\partial_T-c\,\partial_\theta\right)\,u+
\partial^2_\theta\zeta(\theta,T)\right)\nonumber\\
 \end{eqnarray}
where $~~\dot{}\equiv \frac{d}{d T}$ and 
the notation $\xi(x,t)= \zeta(\theta,T)$ was used.
 We seek an asymptotic solution
of the form
\begin{eqnarray}\label{a4}
u=u_0+\epsilon\,u_1+\epsilon^2\,u_2+\cdots
\end{eqnarray}
with 
\begin{eqnarray}\label{a5}
c=c_0+\epsilon\,c_1+\epsilon^2\,c_2+\cdots
\end{eqnarray}
Inserting Eqs (\ref{a4}) and (\ref{a5}) into Eq. (\ref{a3}) and
collecting powers of $\epsilon$ we get

$\epsilon^0$:
\begin{eqnarray}\label{a6}
\partial^2_\theta\left((c^2_0-1)
\,u_0
-c^2_0\,\partial^2_\theta\,u_0-f(u_0)\right)=0,
\end{eqnarray}


$\epsilon^1$:
\begin{eqnarray}\label{a7}
\partial^2_\theta\left((c_0^2-1)
\,u_1
-c^2_0\,\partial^2_\theta\,u_1-\, f'(u_0)\,u_1
\right)=\partial_\theta\,F_1,
\end{eqnarray}

\begin{eqnarray}\label{a71}
F_1&=&-(1-\partial^2_\theta)\left(2\,c_0\,c_1
\,\partial_\theta-2\,c_0\,\partial_T-
\dot{c}_0\right)\,u_0\nonumber\\
&-&\nu_m\,c_0\,\partial_\theta^m\,u_0+
\partial_\theta\zeta(\theta,T) ;
\end{eqnarray}

$\epsilon^2$:
\begin{eqnarray}\label{a8}
& &\partial^2_\theta\left((c_0^2-1)
\,u_2
-c^2_0\,\partial^2_\theta\,u_2-\, f'(u_0)\,u_2
\right)=\partial_\theta\,F_2\nonumber\\
&-&(1-\partial^2_\theta)\partial_T^2\,u_0+\nu_m\,\partial_\theta^{m}\,\left(\partial_T\,u_0 -
c_0\,\partial_\theta\,u_1 -
c_1\,\partial_\theta\,u_0\right),\nonumber\\
\end{eqnarray}

\begin{eqnarray}\label{a81}
F_2&=&-
(1-\partial^2_\theta)\{2\,c_0\,c_1\,\partial_\theta\,u_1 +
 (c_1^2+2 c_0 c_2)\,\partial_\theta\,u_0-\nonumber\\
& & 2\,c_0\,\partial_T\,u_1 - 2
 c_1 \partial_T\,u_0-
\dot{c}_0 \,u_1 - \dot{c}_1\,u_0 \}\,\nonumber\\
& &+\frac{1}{2}\,\partial_\theta\left( f''(u_0)
\,u_1^2\right).
\end{eqnarray}

Integrating twice Eq. (\ref{a6}) under vanishing boundary conditions for
$u_0$ at infinity, we obtain the equation
\begin{eqnarray}\label{a9}
(c^2_0-1)
\,u_0
-c^2_0\,\partial^2_\theta\,u_0-f(u_0)=0.
\end{eqnarray}
In the case of the power-like anharmonic potential 
 \begin{eqnarray}\label{a10}
 U_{anh}=\frac{1}{p}\,u^p, ~~~p>2\end{eqnarray}
 the solution of Eq. (\ref{a9}) has the form
 \begin{eqnarray}\label{a11}
u_0=\left(\frac{p}{2}(c^2_0-1)\right)^{1/(p-2)}\,\sech^{2/(p-2)}\left
(\frac{p-2}{2\ell}\,\theta\right)
\end{eqnarray}
where the parameter $\ell=\frac{c_0}{\sqrt{c^2_0-1}}$ characterizes
the width of the excitation.

Integrating Eq. (\ref{a7}) under vanishing boundary conditions for
$\partial_\theta\,u_1$ at $\theta\rightarrow\pm \infty$  and for $u_1\rightarrow 0$ 
as $\theta\rightarrow \infty$ we get
\begin{eqnarray}\label{a12}
(c_0^2-1)
\,u_1
-c^2_0\,\partial^2_\theta\,u_1-\,f'(u_0)\,u_1= 
-\int\limits_\theta^{\infty}\,F_1(\bar{\theta})\,d \bar{\theta}.\nonumber\\
\end{eqnarray}
The  homogeneous part of Eq. (\ref{a12})
has two linearly independent  solutions
\begin{eqnarray}\label{a13}
v_1=\partial_\theta\,u_0,~~~v_2=v_1\,
\int\limits_{0}^{\theta}
\frac{d \bar{\theta}}{\left(\partial_{\bar{\theta}}\,u_0\right)^2}
\end{eqnarray}
with the Wronskian
$v_2\,\partial_\theta\,v_1-v_1\,\partial_\theta\,v_2=1$. Taking into
account that of these solutions only $v_1$ vanishes as
$\theta\rightarrow\pm\infty$,  the compatibility condition for a bounded
solution of the inhomogeneous equation (\ref{a12}) is
\begin{eqnarray}\label{a14}
\int\limits_{-\infty}^{\infty}\,d\theta\,v_1(\theta)
\int_{\theta}^{\infty}\,\,F_1(\bar{\theta})\,d \bar{\theta}\equiv
\int\limits_{-\infty}^{\infty}\,u_0\,F_1\,d\theta\,=0
\end{eqnarray}
Inserting Eq. (\ref{a71}) into Eq. (\ref{a14}) we obtain that the
compatibility condition in the order $\epsilon^1$ has the form
\begin{eqnarray}\label{a15}
\partial_T\,\left(c_0\,\langle u^2_0+
(\partial_\theta\,u_0)^2\rangle\right)=\nu_m\,c_0\,
\langle u_0\,
\partial^m_\theta u_0\rangle+
\langle u_0\,\partial_\theta\zeta\rangle\nonumber\\
\end{eqnarray}
where the notation
\begin{eqnarray}\label{not}
\langle g \rangle\equiv \int\limits_{-\infty}^{\infty}g(\theta)\,d\theta
\end{eqnarray}
was introduced.
In the same way, we obtain from Eq. (\ref{a8}) that the
compatibility condition in the order $\epsilon^2$ has the form

\begin{eqnarray}\label{a16}
&&\partial_T\,
\bigg[2 c_0\,\langle u_1\,\left(1-\partial^2_\theta\right)\,u_0\rangle+
c_1\,\langle u^2_0+
(\partial_\theta\,u_0)^2\rangle\bigg ]+\nonumber\\
&&\langle u_0(\theta)
\int\limits_{\theta}^{\infty}\,d\bar{\theta}\,
\partial^2_T u_0(\bar{\theta})\rangle + 
\frac{1}{2}(c_0^2-1)\,M^2=\nonumber\\
&&\nu_m\,
\left(2 c_0 \langle u_1\,\partial_\theta^m u_0\rangle + c_1 
\,\langle u_0\,
\partial^m_\theta u_0\rangle -\langle u_0\,\partial_T\,
\partial^{m-1}_\theta u_0\rangle\right)\, .\nonumber\\
\end{eqnarray}
Here \begin{displaymath}
M=\lim_{\theta\rightarrow-\infty}\,u_1(\theta)\end{displaymath} 
and as seen from Eq.
(\ref{a7}) this quantity is determined by the expression 
\begin{eqnarray}\label{a161}
M&=&-\frac{1}{c^2_0-1}\langle F_1(\theta)\rangle\nonumber\\
&=&-\frac{1}{c^2_0-1}\,\bigg(2 c_0\,\partial_T\langle
u_0\rangle+\dot{c}_0\,\langle u_0\rangle-\nu_0\,\delta_{m 0}\,c_0\,
\langle u_0\rangle\bigg.\nonumber\\
&&\bigg.+\zeta(\infty,T)-\zeta(-\infty,T)\bigg)
\end{eqnarray}
Note that in deriving the compatibility condition (\ref{a16}) the relation
\begin{eqnarray}
\frac{1}{2}\langle u_0\,
\partial_\theta \left(u_1^2\,f''(u_0)\right)\rangle=
-\langle u_1\,
\partial_\theta \left(u_1\, f'(u_0)\right)\rangle\end{eqnarray}
and Eqs (\ref{a7})-(\ref{a71}) were used.

Further simplification of the compatibility condition (\ref{a16})
may be achieved by using the relation
\begin{eqnarray}\label{a20}
\left(c^2_0-1-f'(u_0)-c_0^2\,\partial^2_\theta\right)\,
\frac{\partial u_0}{\partial\,c^2_0}=-(1-\partial_\theta^2)\,u_0
\end{eqnarray}
which can be obtained by differentiating Eq. (\ref{a9}) with respect
to $c^2_0$, and the relation
\begin{eqnarray}\label{a21}
\left(1-\frac{1}{c^2_0-1}\,f'(u_0)-\ell^2\,\partial^2_\theta\right)\,
\frac{\partial u_0}{\partial\,\ell^2}=\partial_\theta^2\,u_0
\end{eqnarray}
which can be obtained from the equation
\begin{eqnarray}
\label{a22}
\left(\ell^2\,\partial^2_\theta-1\right)\,u_0+\frac{1}{c^2_0-1}\,f(u_0
)=0
\end{eqnarray} by differentiating it with respect to $\ell^2$.
Thus by using Eqs (\ref{a12}) and (\ref{a20}) we get
\begin{eqnarray}\label{a23}
&&\langle u_1\,(1-\partial_\theta^2)\,u_0\,
\rangle =\frac{1}{2}\, c_1\frac{\partial}{\partial\,c_0}\,\langle
\left(u^2_0+(\partial_\theta u_0)^2\right)\rangle\nonumber\\
&-&\frac{1}{4\,c_0}M\,(c^2_0-1)\,
\frac{\partial\,\langle
u_0\rangle}{\partial\,c_0}\nonumber\\
&-&\frac{1}{2 c_0}\,\left\langle
\frac{\partial u_0}{\partial\,c_0}\left(\zeta(\theta,T)-
\frac{1}{2}(\zeta(-\infty,T)+\zeta(\infty,T))\right)\right\rangle .\nonumber\\
\end{eqnarray}
In the same way by using Eqs (\ref{a12}) and (\ref{a21}) we get
\begin{eqnarray}\label{a24}
\langle u_1\,\partial_\theta^2\,u_0\rangle =\frac{c^2_0-1}{4 c_0^2}\,M\,
\langle u_0\rangle -\frac{c_1}{2 c_0}\langle 
u^2_0-(\partial_\theta u_0)^2\rangle+\nonumber\\
\frac{1}{2 c^2_0}\,
\left\langle \theta\,\partial_{\theta}u_0\,\left(\zeta(\theta,T)-
\frac{1}{2}(\zeta(-\infty,T)+\zeta(\infty,T))\right)\right\rangle.\nonumber\\
\end{eqnarray}
Inserting Eqs (\ref{a23}) and (\ref{a24}) into Eq. (\ref{a16}) we
obtain that the compatibility condition in the order $\epsilon^2$
takes the form
\begin{eqnarray}\label{a25}
&&\frac{\partial}{\partial\,T}\bigg\{\bigg.
c_1\,\frac{\partial}{\partial\,c_0}
\left(c_0\langle u^2_0 + (\partial_\theta u_0)^2\rangle\right)\nonumber\\
&-&\frac{1}{2}\,\left((c^2_0-1)\,M-\dot{c}_0\,\
\langle u_0\rangle\right)
\frac{\partial \langle u_0\rangle}{\partial c_0}
\bigg.\bigg\}+
\frac{1}{2}\,(c_0^2-1)\,M^2\nonumber\\
&-&\frac{1}{2}\,\dot{c}_0^2\,\left(\frac{\partial\,\langle
u_0\rangle}{\partial\,c_0}\right)^2\,-
\dot{c}_0\,\frac{\partial^2}{\partial c^2_0}\bigg\langle\bigg.
u_0\bigg(\bigg.\zeta(\theta,T)\nonumber\\
&-&\frac{1}{2}(\zeta(-\infty,T)-\zeta(\infty,T))\bigg.\bigg)
\bigg.\bigg\rangle=\nu_m\,D_m\nonumber\\
\end{eqnarray}
where the right-hand-side is determined by the expressions
\begin{eqnarray}
\label{a26}
& &D_0=\frac{c^2_0-1}{2\,c_0}\,M\,\left(\langle u_0\rangle-c_0\,
\frac{\partial\,\langle
u_0\rangle}{\partial\,c_0}\right)  +
c_1\,\langle (\partial_\theta u_0)^2\rangle  \nonumber\\
&+& c_1 c_0\,\frac{\partial}{\partial\,c_0}\,\left(\langle
u^2_0\rangle+\,\langle (\partial_\theta u_0)^2\rangle\right)+\frac{\dot{c}_0}{2}\,\langle
u_0\rangle\,\frac{\partial\,\langle u_0\rangle}{\partial\,c_0} \nonumber\\
&+&\bigg\langle\bigg.\left(\frac{\theta}{c_0}\,\partial_{\theta}u_0-
\frac{\partial u_0}{\partial\,c_0}\right)\left(\zeta(\theta,T)-
\frac{1}{2}(\zeta(-\infty,T)\right.\nonumber\\
&+&\bigg.\zeta(\infty,T))\bigg)\bigg.\bigg\rangle,\nonumber\\
&&D_2=\frac{c^2_0-1}{2\,c_0}\,M\,\langle u_0\rangle-c_1\,\langle
u_0^2\rangle\nonumber\\
&+&\frac{1}{c_0}\,\langle\theta\,\partial_{\theta} u_0\,
\left(\zeta(\theta,T)-
\frac{1}{2}(\zeta(-\infty,T)+\zeta(\infty,T))\right)\bigg.\bigg\rangle.\nonumber\\
\end{eqnarray}
To find how the soliton profile changes in the presence of damping
it is convenient to represent the function $u_1$ in the form (see
\cite{grim})
\begin{eqnarray}\label{a17}
u_1=\frac{1}{2} M + w + v
\end{eqnarray}
where $w\rightarrow 0$ as $\theta\rightarrow\pm\infty$ while
$v\rightarrow \mp\frac{1}{2} M$ as $\theta\rightarrow\pm\infty$.
The functions $w$ and $v$ satisfy the equations
\begin{eqnarray}\label{a18}
c^2_0\,\partial^2_\theta w + \left(f'(u_0)
-c^2_0+1\right)\,w=G^w(\theta)
\end{eqnarray}
\begin{eqnarray}\label{a181}
c^2_0\,\partial^2_\theta v + \left(f'(u_0)
-c^2_0+1\right)\,v=G^v(\theta)
\end{eqnarray}
where
\begin{eqnarray}\label{a19}
G^w(\theta)=2 c_0
c_1\,(1-\partial^2_\theta)u_0-\frac{1}{2} M \,f'(u_0) 
\end{eqnarray}
\begin{eqnarray}\label{a191}
G^v(\theta)=-\int\limits_{0}^{\theta}\,\{(1-\partial^2_{\bar{\theta}})
\left(2\,c_0\,\partial_T +\dot{c}_0
\right) - \nu_m\,c_0\,\partial^m_{\bar{\theta}}\}\,u_0(\bar{\theta})\,
d\bar{\theta} \nonumber\\
\end{eqnarray}
Using the functions (\ref{a13}) it is straightforward to see  that 
the solutions of 
Eqs (\ref{a18}) and (\ref{a181})  are determined by the expressions
\begin{eqnarray}\label{a}
w=\frac{1}{c^2_0}\int\limits_{0}^{\theta}\left(v_1(\theta)\,v_2(\bar
{\theta})-v_2(\theta)\,v_1(\bar
{\theta})\right)\,G^w(\bar{\theta})\,d\bar{\theta},\nonumber\\
v=\frac{1}{c^2_0}\int\limits_{0}^{\theta}\left(v_1(\theta)\,v_2(\bar
{\theta})-v_2(\theta)\,v_1(\bar
{\theta})\right)\,G^v(\bar{\theta})\,d\bar{\theta}.
\end{eqnarray}
and $c_1$ to be obtained via Eq. (\ref{a25}).

\section{Discretization of the Bq equation}
The Bq equations reads
\begin{eqnarray}\label{eqcappen}
\partial_t^2 u(x,t)-\partial^2_x u(x,t)-\partial^2_t\partial_x^2 u(x,t)
-\partial_x^2 \left(f(u(x,t))\right)=\nonumber\\
     F(x,t)\quad\quad
\end{eqnarray}
where $F(x,t)$ are external forces and/or dissipation. 
By defining the variable $v(x,t)=\partial_t u(x,t)$
Eq. (\ref{eqcappen}) can be reduced to two partial differential equations of first order
in time, namely
\begin{eqnarray}\label{eqcappen1}
\partial_t v(x,t)&=&\partial^2_x u(x,t)+\partial^2_x\partial_t v(x,t)
+\partial_x^2 \left(f(u(x,t))\right)\nonumber\\
& &+F(x,t)\nonumber\\
\partial_t u(x,t)&=&v(x,t).
\end{eqnarray}
By using finite-difference discretization in the space-domain
Eqs. (\ref{eqcappen1}) take the form
\begin{eqnarray}\label{eqcappen2}
\dot{v}_i(t)&=&\frac{u_{i+1}(t)-2u_{i}(t)+u_{i-1}(t)}{\Delta x^2}\nonumber\\
& &+\frac{\dot{v}_{i+1}(t)-2\dot{v}_{i}(t)+\dot{v}_{i-1}(t)}{\Delta  x^2}\nonumber\\
& &+\frac{f(u_{i+1}(t)))-2f(u_{i}(t))+f(u_{i-1}(t))}{\Delta
x^2}\nonumber\\
& &+ F_i(t),\nonumber\\
\dot{u}_i(t)&=&v_i(t)
\end{eqnarray}
where $\dot{}\equiv \frac{d}{d t}$,
$u_{i}(t)=u(x_i,t)$, $v_{i}(t)=v(x_i,t)$, $f(u_{i}(t))=u^n(x_i,t)$ and
$F_i(t)=F(x_i,t) $ with $n=2,3$. $x_i=i\,\Delta x$ where $\Delta x$ is the 
mesh size of the space variable and $i=1,2,\cdot\cdot\cdot,N$. The  length of the
system $L=N\,\Delta x$. In the numerical integration process we use
periodic boundary conditions, namely $u_{0}(t)=u_{N}(t)$ and
$u_{N+1}(t)=u_{1}(t)$. The same boundaries are used for the variables
$v_{i}(t)$ and $F_i(t)$. If we rewrite Eqs. (\ref{eqcappen2}) so
\begin{eqnarray}\label{eqcappen3}
& &-\dot{v}_{i+1}(t)+(\Delta x^2+2)\dot{v}_i(t)-\dot{v}_{i-1}(t)=\nonumber\\
& &u_{i+1}(t)-2u_{i}(t)+u_{i-1}(t)+\nonumber\\
& &f(u_{i+1}(t))-2f(u_{i}(t))+f(u_{i-1}(t))+\nonumber\\
& &\Delta x^2 \, F_i(t),\\
\label{eqcappen4}
& &\dot{u}_i(t)=v_i(t),
\end{eqnarray}
they can be regarded as a vectorial equations so
\begin{eqnarray}\label{eqcappen5}
\hat{\bf A}\dot{\bf v}&=&{\bf G},\nonumber\\
\dot{\bf u}&=&{\bf v}
\end{eqnarray}
where $\dot{u}_i$ and $\dot{v}_i$ are elements of the vectors $\dot{\bf u}$ and
$\dot{\bf v}$, respectively. The elements $G_i$ of the vector ${\bf G}$ are the
r.h.s. of (\ref{eqcappen3}) and the square matrix
\begin{eqnarray}\label{eqcappen6}
\hat{\bf A}=
\left(
\begin{array}{ccccccccc}
\Delta  & -1     & 0      & 0  & \cdot\cdot\cdot & \cdot\cdot\cdot & 0 & 0 & -1\\
 -1     & \Delta & -1     & 0  & \cdot\cdot\cdot & \cdot\cdot\cdot & 0& 0 & 0\\
 &\cdot\cdot\cdot &\cdot\cdot\cdot &\cdot\cdot\cdot
 &\cdot\cdot\cdot &\cdot\cdot\cdot &\cdot\cdot\cdot &\cdot\cdot\cdot &\\
0 & \cdot\cdot\cdot & 0 &-1 & \Delta & -1 & 0 &  \cdot\cdot\cdot & 0\\
 &\cdot\cdot\cdot &\cdot\cdot\cdot &\cdot\cdot\cdot
 &\cdot\cdot\cdot &\cdot\cdot\cdot &\cdot\cdot\cdot &\cdot\cdot\cdot &\\
0&0&  \cdot\cdot\cdot& \cdot\cdot\cdot& 0  & 0 &-1 & \Delta & -1 \\
-1 & 0& 0& 0&\cdot\cdot\cdot & \cdot\cdot\cdot & 0 &-1 & \Delta  \\
\end{array}
\right)_{N\times N}\nonumber
\end{eqnarray}
with $\Delta =\Delta x^2+2$. Notice that this tridiagonal matrix is
cyclic because we use periodic boundary conditions \cite{numerics}.
From (\ref{eqcappen5}) we can derive
\begin{eqnarray}\label{eqcappen7}
\dot{\bf v}&=&\hat{\bf A}^{-1}{\bf G},\nonumber\\
\dot{\bf u}&=&{\bf v},
\end{eqnarray}
therefore at this stage we can use a classical integrator as for
example the Heun algorithm in order to perform the numerical
integration in time.  

\section{Coefficients \label{coef}}
The soliton solution in the case of cubic anharmonicity and
hydrodynamical damping reads
 \begin{displaymath}
u=u_0+\frac{1}{2} M + w + v \nonumber 
\end{displaymath}
with
\begin{eqnarray}
w&=&\sech^2(\phi)\,(A_1+A_2\, \phi\, \tanh(\phi)) \nonumber\\
v&=&
A_3\,\phi\,\sech^2(\phi)+(A_4+(A_5\,\phi^2+A_6)\sech^2(\phi))\,\tanh(\phi)\nonumber\\
&&+A_7\,Tanh^3(\phi)
\nonumber
\end{eqnarray}
where
\begin{eqnarray}
A_1&=&\frac{3\nu}{5}\frac{(3-c_0^2)\sqrt{c_0^2-1}}{2\,c_0^2-1}+3\,c_1\,c_0\nonumber\\
A_2&=&-\frac{3\nu}{5}\frac{(3-c_0^2)\sqrt{c_0^2-1}}{2\,c_0^2-1}-3\frac{c_1}{c_0}\nonumber
\end{eqnarray}
\begin{displaymath}\nonumber
\begin{array}{ll}
A_3=\frac{2\nu\sqrt{c_o^2-1}}{5(2c_0^2-1)}  & A_4=-\frac{3}{8}(5c_0^2-3)A_3 \\
  &  \\
A_5=-\frac{1}{2c_0^2}A_3 & A_6=-\frac{1}{8}(17-15c_0^2)A_3 \\
  &  \\
A_7= -\frac{1}{8}(5c_0^2-3)A_3 & M=(5c_0^2-3) A_3 \\
\end{array}\nonumber
\end{displaymath}
The velocities $c_0$ and $c_1$ can be determined by Eqs. (\ref{setm2p3}).

\end{document}